\documentclass[aps,pra,twocolumn]{revtex4}%
\usepackage{amsmath}
\usepackage{amsfonts}
\usepackage{amssymb}
\usepackage{graphicx}%
\providecommand{\U}[1]{\protect\rule{.1in}{.1in}}

\begin{document}
\title{Quantum uncertainty relation saturated by the eigenstates of the harmonic oscillator}
\author{A.~Mandilara and N.~J.~Cerf}
\affiliation{Quantum Information and Communication, \'{E}cole Polytechnique de Bruxelles,
CP~165/59, Universit\'{e} Libre de Bruxelles, 1050 Brussels, Belgium}

\begin{abstract}
We re-derive the Schr\"{o}dinger-Robertson uncertainty principle for the
position and momentum of a quantum particle. Our derivation
does not directly employ commutation relations, but works by reduction to an
eigenvalue problem related to the harmonic oscillator, which can then be further exploited 
to find a larger class of constrained uncertainty relations. We derive an uncertainty relation under
the constraint of a fixed degree of Gaussianity and prove that, remarkably,
it is saturated by all eigenstates of the harmonic oscillator.
This goes beyond the common knowledge that the (Gaussian) ground state of the
harmonic oscillator saturates the uncertainty relation.

\end{abstract}
\maketitle

The Heisenberg uncertainty principle \cite{Heisenberg} captures the difference
between classical and quantum states, and sets a limit on the precision of
incompatible quantum measurements. It has been introduced in the early days of
quantum mechanics, but its form has evolved with the understanding and
formulation of quantum physics throughout the years. The first rigorous
mathematical proof of Heisenberg's uncertainty relation for the canonical
operators of position $\hat{x}$ and momentum $\hat{p}$ ($\left[  \hat{x}%
,\hat{p}\right]  =i\hbar$)
\begin{equation}
(\left\langle \hat{x}^{2}\right\rangle -\left\langle \hat{x}\right\rangle
^{2})(\left\langle \hat{p}^{2}\right\rangle -\left\langle \hat{p}\right\rangle
^{2})\geq\hbar^{2}/4, \label{H}%
\end{equation}
is due to Kennard \cite{Kennard} and Weyl \cite{Weyl}, but only pure states
were considered there. The full proof was later derived following different
methodologies \cite{Mandel'shtam,Moyal,Stoler} (see \cite{Man'ko} for more
details), while the properties of the states saturating this inequality were
also progressively unveiled.

The original uncertainty relation~(\ref{H}) only concerned the operators
$\hat{x}$ and $\hat{p}$, but it was generalized to any pair of Hermitian
operators by Schr\"{o}dinger \cite{Schrodinger} and Robertson \cite{Robertson}%
, in the case of pure states. In the same works, the anticommutator of
$\hat{x}$ and $\hat{p}$ was also included in Eq.~(\ref{H}), yielding a
stronger uncertainty relation
\begin{align}
(\left\langle \hat{x}^{2}\right\rangle -\left\langle \hat{x}\right\rangle
^{2})(\left\langle \hat{p}^{2}\right\rangle -\left\langle \hat{p}\right\rangle
^{2}) &  \nonumber\\
-\frac{1}{4}\left(  \left\langle \hat{x}\hat{p}+\hat{p}\hat{x}\right\rangle
-2\left\langle \hat{x}\right\rangle \left\langle \hat{p}\right\rangle \right)
^{2} &  \geq\hbar^{2}/4,\label{RS}%
\end{align}
that bears their name. The first proof of the Schr\"{o}dinger-Robertson (SR)
uncertainty relation for position and momentum, Eq.~(\ref{RS}), in the general
case including mixed states is probably due to Moyal \cite{Moyal}, and, for
any pair of not-necessarily Hermitian operators, to Dodonov, Kurmyshev and
Man'ko \cite{Kurmyshev}. In this latter work, the states of minimum
uncertainty or \textit{minimizing states} (MSs) for the SR inequality were
identified as the pure states with a Gaussian wavefunction. 
Such Gaussian states are ubiquitous in physics as they play a major role for example in quantum optics 
(e.g., coherent states of the light field \cite{Scully}), atomic physics (e.g., collective excitations of an atomic ensemble \cite{Hammerer}),
optomechanics (e.g., nanomechanical oscillators \cite{Aspelmeyer}), supraconductivity (e.g. superconducting LC circuits \cite{Chiorescu}), etc.

In this Rapid Communication, we revisit the status of Gaussian states in the context of uncertainty relations
by exhibiting a connection with the harmonic oscillator and showing that remarkably, {\it all} its eigenstates -- not just its ground state -- appear as minimum uncertainty states. 
We first re-derive the SR inequality and corresponding MSs by using a variational method and standard algebraic tools. 
This derivation reveals the direct link between the quadratic order of Eqs. (\ref{H}) and (\ref{RS}) in $\hat{x}$ and $\hat{p}$ and the fact that we deal with the quadratic Hamiltonian of a harmonic oscillator.
Then, we move on to find \textit{bounded} uncertainty relations \cite{Dodonov}, which give stronger bounds than Eq.~(\ref{RS}) for states on which some {\it a priori }information is
known, such as their purity \cite{Man'ko} or entropy \cite{Bastiaans}. Specifically, we derive a \textit{Gaussianity-bounded} uncertainty relation,
depending on the degree of Gaussianity of the state as measured by a parameter $g$ that we introduce. We identify its corresponding set of MSs and find among them all the eigenstates 
of the harmonic oscillator. This yields a fundamental new set of non-Gaussian minimum-uncertainty states, going beyond the common knowledge on the Heisenberg principle.

Although the uncertainty relations, being at the root of quantum mechanics, have been investigated in various situations, such as multi-dimensional \cite{Sudarshan,Simon2} or mixed states \cite{Bastiaans,Man'ko,Ponomarenko}, our results imply that there is more to gain by analyzing them under the perspective of the Gaussian character of a state. Non-Gaussian states of light can now be handled in the lab \cite{Polzik,Grangier2,Parigi07,Furusawa2,Zavatta09,Bimbard10}
and have been proven essential in the field of continuous-variable quantum information \cite{NoGoPurif1,NoGoPurif2,NoGoPurif3,NoGoCorr,NoGoCommit}, but they remain hard to classify. Identifying states of minimum uncertainty among them may lead to a better understanding of the structure of the state space in infinite dimension and, since the Heisenberg principle is at the heart of the limitations on measurement precision \cite{Braunstein}, to the possible development of novel concepts in quantum metrology.

\textit{Unconstrained SR relation}.---We introduce our method as a way
to find the MSs of the SR uncertainty relation. Consider a quantum
state characterized by its density operator $\hat{\rho}$. Its covariance
matrix is defined as
\begin{equation}
\gamma=\left(   \begin{array}{cc}
\left\langle \hat{x}^{2}\right\rangle -\left\langle \hat{x}\right\rangle^{2} & 
\frac{1}{2} \left\langle \hat{x}\hat{p}+\hat{p}\hat{x}\right\rangle -\left\langle \hat{x}\right\rangle \left\langle \hat{p}\right\rangle   \\
\frac{1}{2} \left\langle \hat{x}\hat{p}+\hat{p}\hat{x}\right\rangle -\left\langle \hat{x}\right\rangle \left\langle \hat{p}\right\rangle   
& \left\langle \hat{p}^{2}\right\rangle -\left\langle \hat{p}\right\rangle^{2}
\end{array}   \right)
\label{Co}
\end{equation}
where $\left\langle \cdot \right\rangle  = \mathrm{Tr}(\hat{\rho} \; \cdot)$ stand for quantum expectation values in state $\hat{\rho}$.
Hereafter, we define the \textit{uncertainty} of the state $\hat{\rho}$ as the dimensionless variable  
\begin{equation}
\alpha=2 \, (\det\mathbf{\gamma})^{1/2}  / \hbar.
\label{alpha}
\end{equation}
which is  simply the square root of the left-hand side of Eq.~(\ref{RS}) divided by $\hbar/2$. 
We write it with the determinant of $\mathbf{\gamma}$ to emphasize that it remains invariant under any linear canonical transformation in $x$ and $p$, that is, under any operation modeled by a Hamiltonian that is quadratic in $\hat{x}$ and $\hat{p}$ (see Appendix~A). This brings the strong simplification that it is sufficient to confine our search for MSs among states that satisfy the constraints
\begin{align}
\mathrm{Tr}\left(  \hat{\rho}\hat{x}\right)   &  =\mathrm{Tr}\left(  \hat
{\rho}\hat{p}\right)  =0\label{c1}\\
 \mathrm{Tr}\left(  \hat{\rho}\left(  \hat{x}\hat{p}+\hat{p}\hat{x}\right)
\right)  &=  \mathrm{Tr}\left(  \hat{\rho}\left(  \hat{x}^{2}-\hat{p}%
^{2}\right)  \right)  =0\label{c2}%
\end{align}
that is, whose mean values vanish and $\gamma$ is proportional to the identity.
Under these conditions, Eq.~(\ref{alpha}) can be expressed as
\begin{equation}
\alpha=\mathrm{Tr}\left(  \hat{\rho}\left(  1+ 2\hat{n} \right)  \right) , \label{un}
\end{equation}
where $\hat{n}=(\hat{x}^2+\hat{p}^2-1)/2$ is the number operator for the harmonic oscillator (from now on, we assume $\hbar=1$ and take the particle's mass and angular frequency equal to one).

Now, let us proceed with the minimization of $\alpha$ under constraints (\ref{c1}) and (\ref{c2}) by using the Lagrange multipliers method. For any density operator $\hat{\rho}$, there is an eigenbasis $\left\{  \left\vert \Psi_{n}\right\rangle \right\} $ such that $\hat{\rho}=\sum c_{n}\left\vert \Psi
_{n}\right\rangle \left\langle \Psi_{n}\right\vert $ with $0\leq c_{n}\leq1$
and $\sum c_{n}=1$. It is more convenient to define the
unnormalized vectors $\left\vert \psi_{n}\right\rangle =\sqrt{c_{n}}\left\vert
\Psi_{n}\right\rangle $ and rewrite the state as
$\hat{\rho}=\sum_{n}\left\vert \psi_{n}\right\rangle \left\langle \psi
_{n}\right\vert$, while imposing the additional constraint
\begin{equation}
\mathrm{Tr}\left(  \hat{\rho}\right)  =1.\label{n1}%
\end{equation}
Then, choosing an orthonormal basis $\left\{  \left\vert i\right\rangle
\right\}  $ to decompose the vectors $\left\vert \psi_{n}\right\rangle
=\sum\psi_{n}^{i}\left\vert i\right\rangle $, we can re-express
the uncertainty (\ref{un}) and constraints (\ref{c1}), (\ref{c2}), and
(\ref{n1}) as functions of the $\psi_{n}^{i}$'s.  We define the Lagrange multipliers $\lambda_{k}^{\prime}$ and
consider the ``uncertainty'' functional
\begin{align}
\tilde{\alpha} &  =\alpha+\lambda_{1}^{\prime}\mathrm{Tr}\left(  \hat{\rho
}\right)  +\lambda_{2}^{\prime}\mathrm{Tr}\left(  \hat{\rho}\hat{x}\right)
+\lambda_{3}^{\prime}\mathrm{Tr}\left(  \hat{\rho}\hat{p}\right)  \nonumber\\
&  +\lambda_{4}^{\prime}\mathrm{Tr}\left(  \hat{\rho}\left(  \hat{x}\hat
{p}+\hat{p}\hat{x}\right)  \right)  +\lambda_{5}^{\prime}\mathrm{Tr}\left(
\hat{\rho}\left(  \hat{x}^{2}-\hat{p}^{2}\right)  \right) ,
\label{functio}
\end{align}
which implicitely depends on the complex amplitudes $\psi_{n}^{i}$'s.
Extremizing $\tilde{\alpha}$ yields conditions on these amplitudes (see Appendix~A), which 
read as conditions on the unormalized eigenvectors $\left\vert \psi_{n}\right\rangle $ defining the minimizing
state $\hat{\rho}$, namely
\begin{align}
[ \hat{n}+1/2+\lambda_{1}+\lambda_{2}\hat{x}+
\lambda_{3}\hat{p} +\lambda_{4}  (\hat{x}\hat{p}+\hat{p}\hat{x}) 
&  \nonumber\\
 + \lambda_{5}\left(  \hat{x}^{2}-\hat{p}^{2}\right)  ]   \left\vert \psi_{n}\right\rangle  &  =0\label{Hami}%
\end{align}
where  $\lambda_{k}=2\lambda_{k}^{\prime}$. Introducing the Hermitian operator
\begin{equation}
\hat{H}=\hat{n}+\frac{1}{2}+\lambda_{2}\hat{x}+\lambda_{3}\hat{p}+\lambda_{4}\left(
\hat{x}\hat{p}+\hat{p}\hat{x}\right)  +\lambda_{5}\left(  \hat{x}^{2}-\hat
{p}^{2}\right)  ,\label{Hamo}%
\end{equation}
we can rewrite Eq.~(\ref{Hami}) as $\hat{H}\left\vert \psi_{n}\right\rangle
=-\lambda_{1}\left\vert \psi_{n}\right\rangle $, $\forall n$, leading to the
necessary condition that the eigenvectors $\left\vert \psi_{n} \right\rangle$ defining the MSs must be degenerate eigenvectors of $\hat{H}$
corresponding all to the same eigenvalue.

Thus, one should diagonalize $\hat{H}$ in order to proceed with the identification of the MSs.
As explained in  Appendix~A, there exists a linear canonical transformation in $x$ and $p$ that transforms 
$\hat{H}$ onto the Hamiltonian of the harmonic oscillator, $\hat{H}_{0}=\hat{n}+1/2$.
Obviously, this means that $\hat{H}$ has the same eigenvalues as $\hat{H}_{0}$ and that its eigenvectors are $U\left\vert n\right\rangle$, where $\left\vert n\right\rangle$ are the number states (eigenstates of $\hat{H}_{0}$) and $U$ is the Gaussian unitary corresponding to this canonical transformation.
Since $\hat{H}_{0}$ does not possess any degeneracy in its spectrum, the same holds for $\hat{H}$
and therefore the only possibility is that the MSs is a pure state of the type
$\hat{\rho}=U\left\vert n\right\rangle \left\langle n\right\vert U^{\dag}$.
Among these states, we must keep those satisfying constraints~(\ref{c1})-(\ref{c2}), which are the number states
$\left\vert n\right\rangle $, so that the state that minimizes the
uncertainty (\ref{un}) is obviously the ground state $\left\vert 0\right\rangle $. Of course, by plugging $\hat{\rho}=\left\vert
0\right\rangle \left\langle 0\right\vert $ into Eq.~(\ref{un}), we recover the lower bound of the SR relation, $\hbar^{2}/4$. 
By acting with linear canonical transformations on $\left\vert 0\right\rangle$, 
we obtain all Gaussian pure states, which is the well-known set of MSs for the SR uncertainty relation (see Appendix~A).
This was a long detour to re-derive Eq.~(\ref{RS}), but this connection with the harmonic oscillator turns out to be crucial in what follows.

\textit{Degree of Gaussianity}.---Our method works by reduction to a constrained optimization
problem (even for solving the unconstrained SR inequality), so it can be
simply adapted to find the MSs with an extra constraint on Gaussianity. 
Several measures of non-Gaussianity have been used in
the literature \cite{Dodonov,Bastiaans,Genoni,Genoni2,Simon}, but we
instead suggest using a parameter $g$ capturing the degree of Gaussianity,
inspired from our former work on non-Gaussian states with positive Wigner function \cite{Hudson,Positivity}. 
Denoting as $\hat{\rho}_{G}$ the Gaussian state that has the same covariance matrix $\gamma$ (and same mean values $\langle \hat{x} \rangle$ and $\langle \hat{p} \rangle$) as state $\hat{\rho}$, 
we define the Gaussianity of $\hat{\rho}$ as
\begin{equation}
g=\mathrm{Tr}\left(  \hat{\rho}\hat{\rho}_{G}\right)  /\mathrm{Tr}\left(
\hat{\rho}_{G}^{2}\right)  .\label{g}%
\end{equation}
It is more appropriate for our purposes and also has merits on its own, see Appendix~B.
Its main properties are as follows (see  Appendix~B for the proofs):
\textit{(i)} $g$ is invariant under linear canonical transformations in $x$ and $p$;
\textit{(ii)} $g$ is a bounded quantity, that is, $2/e\leq g\leq2$, and $g=1$ for Gaussian states (but the converse is not true);
\textit{(iii)} $g$ provides a necessary criterion for the strict positivity of the
Wigner function of a state.

Let us briefly address possible experimental means for a direct estimation of $g$ without going through a full
state tomography procedure of $\hat{\rho}$ (see Appendix~B). The trace overlap $\mathrm{Tr}\left(\hat{\rho}\hat{\rho}_{G}\right)$
can be estimated  using the  eight-port homodyne detection
scheme that is usually employed to estimate the $Q$-function of a state in quantum optics 
by simultaneously measuring the $x$ and $p$ quadratures of the two output modes of a balanced beam splitter \cite{Scully}. 
If we inject the state $\hat{\rho}$ together with $\hat{\rho}_{G}$ (instead of the vacuum state $\left\vert 0\right\rangle $) 
in this beam splitter preceding the quadrature measurements, then the value of the modified $Q$-function at the origin would read
$Q\left(  0,0\right)  =\mathrm{Tr}\left(\hat{\rho}\hat{\rho}_{G}\right) $. This method for measuring $g$ would, however, require first
performing homodyne measurements on $\hat{\rho}$ in order to obtain its covariance matrix $\gamma$ and prepare the Gaussian state $\hat{\rho}_{G}$.

\textit{Gaussianity-bounded SR relation}.---Our optimization method provides a necessary
condition on the extremal solutions since it relies on the Lagrange multipliers method, so that 
concluding on a solution may become complicated if the eigenvectors of $\hat{H}$ 
cannot be identified analytically. Fortunately, finding an uncertainty relation under a fixed-$g$ constraint 
leads to an eigenvalue problem that is analytically solvable.
As before, we can confine our search on states satisfying constraints (\ref{c1}) and (\ref{c2})
since neither $\alpha$ nor $g$ change under linear canonical transformations. For these
states, the corresponding Gaussian state $\hat{\rho}_{G}$ can be expressed
simply as
\begin{equation}
\hat{\rho}_{G}=e^{-\beta\hat{n}}/N\label{symG}
\end{equation}
where $\mathrm{e}^{-\beta}=\frac{\alpha-1}{\alpha+1}$ and $N=(\alpha+1)/2$.
In addition, $\mathrm{Tr}\left(  \hat{\rho}_{G}^{2}\right)  =1/\alpha$.
Instead of minimizing the uncertainty $\alpha$ for a fixed Gaussianity $g$, it is easier
to fix $\alpha$ and search for states of extremal $g$, or extremal overlap
$o=\mathrm{Tr}\left(  \hat{\rho}e^{-\beta\hat{n}}\right)  $. 
The variational procedure for deriving the MSs is completely
analogous, but we extremize $o$ using a
constraint on the uncertainty~(\ref{un}) in addition to Eqs.~(\ref{c1}%
)-(\ref{c2}) and (\ref{n1}). Thus, the ``overlap'' functional is
\begin{align}
\tilde{o} &  =o+\lambda_{1}^{\prime}\mathrm{Tr}\left(  \hat{\rho}\right)
+\lambda_{2}^{\prime}\mathrm{Tr}\left(  \hat{\rho}\hat{x}\right)  +\lambda
_{3}^{\prime}\mathrm{Tr}\left(  \hat{\rho}\hat{p}\right)  \nonumber \\
&  +\lambda_{4}^{\prime}\mathrm{Tr}\left(  \hat{\rho}\left(  \hat{x}\hat{p}+\hat{p}%
\hat{x}\right)  \right)  +\lambda_{5}^{\prime}\mathrm{Tr}\left(  \hat{\rho}\left(
\hat{x}^{2}-\hat{p}^{2}\right)  \right)  \nonumber \\
&  +\lambda_{6}^{\prime}\mathrm{Tr}\left(  \hat{\rho}\left(  2\hat
{n}+1\right)  \right)  \label{Lang}%
\end{align}
depending on six Lagrange multipliers $\lambda_{k}^{\prime}$.
The extremization conditions on the unnormalized eigenvectors $\left\vert \psi_{n}\right\rangle $ of the solution state
can be written as
\begin{align}
[  e^{-\beta\hat{n}}+1/2 + & \lambda_{1}+\lambda_{2}%
\hat{x}+\lambda_{3}\hat{p}+\lambda_{4}\left(  \hat{x}\hat{p}+\hat{p}\hat
{x}\right)     &  \nonumber\\
&  \phantom{\mathrm{\hat{I}}}+\lambda_{5}\left(  \hat{x}^{2}-\hat{p}%
^{2}\right)  +\lambda_{6}\hat{n} ]  \left\vert \psi
_{n}\right\rangle   =0.\label{conn}%
\end{align}
where $\lambda_{k}=2\lambda_{k}^{\prime}$. By defining the Hermitian operator%
\begin{align}
\hat{H}_{1}   =  & e^{-\beta\hat{n}}+\lambda_{2}\hat{x}+\lambda_{3}\hat
{p}+\lambda_{4}\left(  \hat{x}\hat{p}+\hat{p}\hat{x}\right)  \nonumber\\
&  +\lambda_{5}\left(  \hat{x}^{2}-\hat{p}^{2}\right)  +\lambda_{6}\hat
{n}\label{H1}%
\end{align}
we conclude that $\left\vert \psi_{n}\right\rangle $ should be degenerate
eigenvectors of $\hat{H}_{1}$. It can be shown that, without loss of
generality, we can restrict ourselves to states $\left\vert \psi_{n}
\right\rangle =\psi_{n}\left\vert n\right\rangle $, with $\psi_{n}$ being
complex amplitudes (see Appendix~B). For these states, the constraints~(\ref{c1}%
)-(\ref{c2}) are satisfied, and $\hat{H}_{1}$ is replaced by
\begin{equation}
\hat{H}_{2}=e^{-\beta\hat{n}}+\lambda_{6}\hat{n}.\label{H2}%
\end{equation}
The eigenvectors of $\hat{H}_{2}$ are the number states $\left\vert
n\right\rangle $, but, unlike for the harmonic oscillator, double degeneracies
are possible if $\lambda_{6}<0$. As a result, we look for mixtures of two
number states $\hat{\rho}=\left\vert \psi_{n}\right\vert ^{2} \left\vert n\right\rangle \left\langle n \right\vert 
+\left\vert \psi_{m}\right\vert ^{2} \left\vert m \right\rangle \left\langle m \right\vert $
satisfying the normalization constraint $\left\vert \psi_{n}\right\vert^{2}+\left\vert \psi_{m}\right\vert ^{2} =1 $
and uncertainty constraint
\begin{align}
\left\vert \psi_{n}\right\vert ^{2}\left(  2n+1\right)  +\left\vert \psi_{m}\right\vert ^{2}\left(  2m+1\right)   &  =\alpha ,    \label{rr4}%
\end{align}
that achieve the minimum or maximum
\begin{equation}
g=\left\vert \psi_{n}\right\vert ^{2}\frac{2\alpha\left( \alpha-1\right)^{n}}{\left( \alpha+1\right)^{n+1}}
+\left\vert \psi_{m}\right\vert^{2}\frac{2\alpha\left( \alpha-1\right)^{m}}{\left( \alpha+1\right)^{m+1}}. \label{rr2}%
\end{equation}
By supervision, one can see that the minimum $g$ (corresponding to positive
eigenvalues of $\hat{H}_{2}$) is achieved by mixtures of two successive number
states
\begin{equation}
\hat{\rho}_{\min}=r\left\vert n\right\rangle \left\langle n\right\vert
+\left(  1-r\right)  \left\vert n+1\right\rangle \left\langle n+1\right\vert ,
\label{romin}%
\end{equation}
with the parameters $n$ and $r\in[0, 1[$ depending on $\alpha$. The number
states $\left\vert n\right\rangle $ are naturally included in the set for
$r=0$. The maximum $g$ (corresponding to negative eigenvalues of $\hat{H}_{2}%
$) is achieved by mixtures
\begin{equation}
\hat{\rho}_{\max}=r\left\vert 0\right\rangle \left\langle 0\right\vert
+\left(  1-r\right)  \left\vert n\right\rangle \left\langle n\right\vert
\label{romax}%
\end{equation}
in the limit $n\rightarrow\infty$, $r\rightarrow1$, while $\alpha=r+(1-r)(2n+1)$ is kept finite.

\begin{figure}[t]
{\centering{\includegraphics*[width=0.45\textwidth] {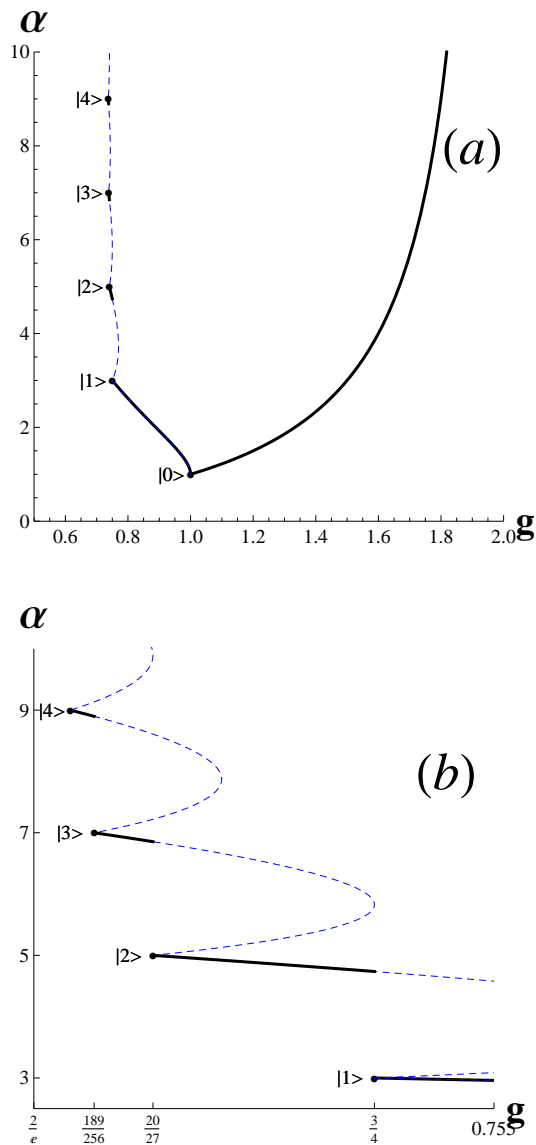}}%
}\caption{(a) Extremal values of the degree of Gaussianity $g$ for a fixed uncertainty $\alpha$
shown as a dashed (blue) line. It is achieved by states
$\hat{\rho}_{\min}$ for $g\leq1$, and $\hat{\rho}_{\max}$ for $g>1$. The line
connecting subsequent number states $\left\vert n\right\rangle $ and $\left\vert
n+1\right\rangle $ is realized by mixtures of them. The Gaussianity-bounded
SR relation corresponds to the part of this extremal line shown as a
solid (black) line. The uncertainty $\alpha$ must have a
value larger or equal to the solid line for a given  $g$. (b) Magnified view of figure (a), where the
discontinuity of the uncertainty relation becomes evident.}%
\label{Fig1}%
\end{figure}

In Figure~\ref{Fig1}, we plot as a dashed line the two extremal values of $g$
for a fixed $\alpha$ as realized by the states of Eqs.~(\ref{romin}%
)-(\ref{romax}). The MSs (i.e., the states minimizing the uncertainty $\alpha$
for fixed degree of Gaussianity $g$) correspond only to some part of this
line, which we show as a solid line. For $g>1$, the situation is simple
and all states $\hat{\rho}_{\max}$ are MSs. For this region, the minimum on the uncertainty as a function of $g$ can be easily derived \begin{equation}\alpha \geq \frac{g}{2-g},\qquad\mathrm{if~}g>1 ,\end{equation} employing Eq.(21) (see Appendix~D). In contrast, for $\frac
{2}{\mathrm{e}}<g\leq1$, the minimum $\alpha$ for fixed $g$ displays
discontinuities. In the interval
\begin{equation}
\frac{\left(  n+1\right)^{n+1}\left(  3+2n\right)  }{\left(  2+n\right)^{2+n}}
<g\leq
\frac{n^{n}\left( 1+2n\right)  }{\left( 1+n\right)^{1+n}%
}\label{inter}%
\end{equation}
$\alpha$ is minimized by the states $\hat{\rho}_{\min}$ with the value of $n$ satisfying Eq. (\ref{inter}).
In particular, we see that all number states $\left\vert n\right\rangle $ are
included in this set for specific values of $g$ corresponding to the upper
bound of Eq.~(\ref{inter}) for different $n$'s. Moreover, in Fig.~\ref{Fig1}(b), we see that most of the MSs (solid line) 
consist of these states $\left\vert n\right\rangle $ and their close neighborhood. Thus, as advertised, 
we conclude that all eigenstates of the harmonic oscillator are extremal 
in the sense that they exhibit the lowest allowed uncertainty given their non-Gaussian character.

For other values of $g$, once the value of $n$ is identified from Eq.~(\ref{inter}), one has to solve the polynomial equation
\begin{equation}
4\alpha\frac{\left( \alpha-1\right)^{n}}{\left( \alpha+1 \right)^{2+n}}\left( 1+n\right)  =g\label{poly}%
\end{equation}
for $\alpha$ in order to find the dependence of the uncertainty $\alpha$ on the Gaussianity $g$.
For $3/4 < g \leq 1$, which covers most of the interesting region, the explicit expression (see Appendix~C) is \begin{equation} \alpha \geq \frac{2+2\sqrt{1-g}-g}{g}, \qquad\mathrm{if~}3/4 < g \leq 1. \end{equation}
In the Appendix~B we  show that, in addition to the states $\hat{\rho}_{\min}$ and $\hat{\rho}_{\max}$, which are phase invariant, the set of MSs comprises all states
with a covariance matrix proportional to the identity which  can be
transformed onto them by phase averaging.  Furthermore, all
states connected to $\hat{\rho}_{\min}$ and $\hat{\rho}_{\max}$ by linear
canonical transformations are MSs as well, since the uncertainty $\alpha$ and Gaussianity $g$ are invariants of the group.

\textit{Conclusions}.---We have exhibited a variational method to derive the Schr\"{o}dinger-Robertson 
uncertainty relation by casting it as  an eigenvalue problem related to the harmonic oscillator.
It follows an ``inverse path'' to the common procedure where the lower bound 
on the uncertainty is derived based on commutators, and the MSs are then identified. 
Such an inverse procedure was put forward by Dodonov and Man'ko for the derivation of
purity-bounded uncertainty relations \cite{Man'ko,Dodonov}, but it appears that our method
is more generally applicable because it is based on the amplitudes of the eigenstates of the MSs
instead of its density matrix elements, see Appendix~D. It is especially useful when constraints are 
included that account for some knowledge on the state. 

In particular, we have found a new uncertainty relation that is
bounded by the degree of Gaussianity $g$ of the state. The state with the overall lowest uncertainty 
 $\alpha$ is of course the Gaussian ground state $\left\vert 0 \right\rangle $ of the harmonic oscillator ($g=1$), 
but we have thus proven that the non-Gaussian states with the lowest uncertainty $\alpha$ for a fixed $g< 1$ include 
as well mixtures of subsequent number states $\left\vert n\right\rangle $ of the harmonic oscillator.
Among these MSs, the number states play a prominant role as they
are the only phase-invariant pure states. We have proven that the number states
are also \textit{extremal} in this uncertainty-related sense, thereby
extending to all (non-Gaussian) eigenstates of the harmonic oscillator the
celebrated minimum-uncertainty property of its (Gaussian) ground state.
Given the considerable attention that non-Gaussian states are attracting in
continuous-variable quantum information theory, see e.g. \cite{Cerf},
unveiling this extremality property of harmonic oscillator states may contribute 
to further fundamental progress in the field, especially in relation with quantum metrology.


AM gratefully acknowledges financial support from the F.R.S-FNRS. This work
was also carried out with the financial support of the F.R.S-FNRS via project
HIPERCOM and the support of the Belgian Federal program IUAP via project Photonics@be.

\appendix
\section{Full proof of the SR uncertainty relation}

Consider an arbitrary quantum state defined by the density operator $\hat{\rho}$. Using standard notations in quantum optics, the covariance matrix $\mathbf{\gamma}$ of this state is  \begin{equation} \gamma_{ij}=\frac{1}{2}\;\mathrm{Tr}(\{(\hat{r}_{i}-d_{i}),(\hat{r}_{j}-d_{j})\}\hat{\rho}) \label{Co2} \end{equation} where $\hat{\mathbf{r}}=(\hat{x},\hat{p})^{T}$ is the vector of position and momentum observables, $\mathbf{d}=$\textrm{$Tr$}$(\hat{\mathbf{r}}\hat{\rho})$ is the displacement vector, and $\{\cdot,\cdot\}$ stands for the anticommutator. We start from the definition of the dimensionless uncertainty \begin{equation} \alpha=\frac{(\det\mathbf{\gamma})^{1/2}}{\hbar/2}.\end{equation} The determinant of $\mathbf{\gamma}$, and consequently the uncertainty $\alpha$, remains invariant under the action of the linear canonical group, the semidirect product of the special linear group $Sp(2,R)$ with thetranslation group $T\left( 2\right) $. In quantum optics, these correspond to Gaussian operations, combining displacements and symplectic transformations \cite{Sudarshan}. Therefore, without loss of generality, we may confine our search of MSs among states with a covariance matrix $\mathbf{\gamma}$ proportional to the identity (the so-called Williamson normal form) and $\mathbf{d=0}$. In other words, we may search for states $\hat{\rho}$ that minimize the uncertainty $\alpha$ while satisfying the constraints \begin{align}\mathrm{Tr}\left(\hat{\rho}\hat{x}\right)&=\mathrm{Tr}\left(\hat{\rho}\hat{p}\right)=0 \label{c12}\\  \mathrm{Tr}\left(\hat{\rho}\left( \hat{x}\hat{p}+\hat{p}\hat{x}\right)\right)& =\mathrm{Tr}\left(\hat{\rho}\left(\hat{x}^{2}-\hat{p}^{2}\right)\right)=0 \label{c22} \end{align} Under these conditions, the uncertainty $\alpha$ can be expressed as \begin{equation} \alpha=\mathrm{Tr}\left(  \hat{\rho}\left(  2\hat{n}+1\right)  \right)  ,\label{un2} \end{equation} where $\hat{n}=\hat{a}^{\dag}\hat{a}$ the number operator.

Now, let us provide more details on the optimization of $\alpha$ under the constraints of Eqs.~(\ref{c12})-(\ref{c22}). It is known that for every density matrix $\hat{\rho}$, a unique eigenbasis $\left\{  \left\vert \Psi_{n}\right\rangle \right\}  $ exists such that $\hat{\rho}=\sum c_{n}\left\vert \Psi_{n}\right\rangle \left\langle \Psi_{n}\right\vert $ with $0\leq c_{n}\leq1$ and $\sum c_{n}=1$. It is more convenient for our purposes to define the unnormalized vectors $\left\vert \psi_{n}\right\rangle =\sqrt{c_{n}}\left\vert \Psi_{n}\right\rangle $, and rewrite $\hat{\rho}$ as \begin{equation}\hat{\rho}=\sum_{n}\left\vert\psi_{n}\right\rangle \left\langle \psi_{n}\right\vert \label{ro}\end{equation} while imposing the additional constraint \begin{equation}\mathrm{Tr}\left(  \hat{\rho}\right)  =1. \label{n12}\end{equation} Then, we choose an orthonormal basis $\left\{  \left\vert i\right\rangle\right\}  $ to decompose the vectors $\left\vert \psi_{n}\right\rangle=\sum\psi_{n}^{i}\left\vert i\right\rangle $ and accordingly re-express the uncertainty (\ref{un2}) as \begin{equation}\alpha=\sum_{n,i,j}\psi_{n}^{i\ast}\psi_{n}^{j}\left\langle i\right\vert\left(  2\hat{n}+1\right)  \left\vert j\right\rangle \label{a1}\end{equation} and the constraints (\ref{c12}), (\ref{c22}), and (\ref{n12}) as \begin{align}\sum_{n,i,j}\psi_{n}^{i\ast}\psi_{n}^{j}\left\langle i\right\vert \hat{x}\left\vert j\right\rangle  &  =0\label{c1b}\\ \sum_{n,i,j}\psi_{n}^{i\ast}\psi_{n}^{j}\left\langle i\right\vert \hat{p}\left\vert j\right\rangle  &  =0\label{c1bb}\\\sum_{n,i,j}\psi_{n}^{i\ast}\psi_{n}^{j}\left\langle i\right\vert \left(\hat{x}\hat{p}+\hat{p}\hat{x}\right)  \left\vert j\right\rangle  &=0\label{q2b}\\\sum_{n,i,j}\psi_{n}^{i\ast}\psi_{n}^{j}\left\langle i\right\vert \left(\hat{x}^{2}-\hat{p}^{2}\right)  \left\vert j\right\rangle  & =0\label{q2bb}\\\sum_{n,i}\psi_{n}^{i}\psi_{n}^{i\ast}  &  =1. \label{n1b}\end{align}

We define the Lagrange multipliers $\lambda_{k}^{\prime}$ and consider the \textquotedblleft uncertainty\textquotedblright\ functional\begin{align}\tilde{\alpha}  &  =\alpha+\lambda_{1}^{\prime}\mathrm{Tr}\left(  \hat{\rho}\right)  +\lambda_{2}^{\prime}\mathrm{Tr}\left(  \hat{\rho}\hat{x}\right)+\lambda_{3}^{\prime}\mathrm{Tr}\left(  \hat{\rho}\hat{p}\right) \nonumber\\&  +\lambda_{4}^{\prime}\mathrm{Tr}\left(  \hat{\rho}\left( \hat{x}\hat{p}+\hat{p}\hat{x}\right)  \right)  +\lambda_{5}^{\prime}\mathrm{Tr}\left(\hat{\rho}\left(  \hat{x}^{2}-\hat{p}^{2}\right)  \right)  , \label{functio2}\end{align} which implicitely depends on the complex amplitudes $\psi_{n}^{i}$'s via the relations Eqs.(\ref{c1b})-(\ref{n1b}). Extremizing $\tilde{\alpha}$ yields conditions on these amplitudes which read as conditions on the unnormalized eigenvectors $\left\vert \psi_{n}\right\rangle $ defining the minimizing state $\hat{\rho}$, namely \begin{align}\lbrack\hat{n}+1/2+\lambda_{1}+\lambda_{2}\hat{x}+\lambda_{3}\hat{p}+\lambda_{4}(\hat{x}\hat{p}+\hat{p}\hat{x}) & \nonumber\\+\lambda_{5}\left(  \hat{x}^{2}-\hat{p}^{2}\right)  ]\left\vert \psi_{n}\right\rangle  &  =0 \label{Hami2}\end{align} where $\lambda_{k}=2\lambda_{k}^{\prime}$. Introducing the Hermitian operator \begin{equation}\hat{H}=\hat{n}+\frac{1}{2}+\lambda_{2}\hat{x}+\lambda_{3}\hat{p}+\lambda _{4}\left(  \hat{x}\hat{p}+\hat{p}\hat{x}\right)  +\lambda_{5}\left(  \hat{x}^{2}-\hat{p}^{2}\right)  , \label{Hamo2}\end{equation} we can rewrite Eq.~(\ref{Hami2}) as $\hat{H}\left\vert \psi_{n}\right\rangle=-\lambda_{1}\left\vert \psi_{n}\right\rangle $, $\forall n$, leading to the necessary condition that the eigenvectors $\left\vert \psi_{n}\right\rangle $ defining the MSs must be degenerate eigenvectors of $\hat{H}$ corresponding all to the same eigenvalue.

One should identify the eigenvectors and eigenvalues of $\hat{H}$. In the language of quantum optics, this is an easy task because one can always apply a squeezing and displacement operation that transforms $\hat{H}$ onto the Hamiltonian of the harmonic oscillator, $\hat{H}_{0}=\hat{n}+1/2$. More precisely, there exists a linear canonical transformation $U=\exp\left( \gamma a^{\dag}-\gamma^{\ast}a\right)  \exp\left(  \beta a^{\dag2}-\beta^{\ast}a^{2}\right)  $ such that $\hat{H}=U\hat{H}_{0}U^{\dag}$, up to a real constant. Obviously, this means that $\hat{H}$ has the same eigenvalues as $\hat{H}_{0}$, up to this constant, and that its eigenvectors are the accordingly transformed number states, $U\left\vert n\right\rangle $, remembering that the number states $\left\vert n\right\rangle $ are the eigenstates of $\hat{H}_{0}$. Among these states, we must keep those satisfying constraints~(\ref{c12})-(\ref{c22}), which are the number states $\left\vert n\right\rangle $, so that the state that minimizes the uncertainty (\ref{un2}) is obviously the vacuum state $\left\vert 0\right\rangle $.

Clearly the vacuum state $\left\vert 0\right\rangle $ is the minimizing state just for the specific class of states that we have chosen for our proof i.e. the states with covariance matrix proportional to identity. The whole set of MSs can be obtained by acting the linear canonical transformations on the $\left\vert 0\right\rangle $ since our proof is invariant under the action of this group. In other words, the ``orbit'' of states that are connected to $\left\vert 0\right\rangle $ by linear canonical transformations (that is, all Gaussian pure states) coincide with the well-known set of MSs for the SR uncertainty relation.

\section{Degree of Gaussianity}

In our work, we have introduced the quantity \begin{equation}g=\mathrm{Tr}\left(  \hat{\rho}\hat{\rho}_{G}\right)  /\mathrm{Tr}\left(\hat{\rho}_{G}^{2}\right) \label{g2}\end{equation} in order to characterize the degree of Gaussianity of a state $\hat{\rho}$. In Eq.~(\ref{g2}), $\hat{\rho}_{G}$ is the reference Gaussian state of $\hat{\rho}$ in the sense that it possess the same covariance matrix $\mathbf{\gamma}$ as $\hat{\rho}$. Here, we exhibit the proofs of the mathematical properties of $g$ that we list in the main text and elaborate on the physical intuition behind its definition.

\textit{(i)} The Gaussianity is invariant under linear canonical transformations, i.e., symplectic transformations $Sp(2,R)$ and translations $T(2)$.

\textit{Proof.} A Gaussian operation $U_{G}$ acting on a state $\hat{\rho}$ can be always translated as a sequence of displacement, rotation, and squeezing of the Wigner function of $\hat{\rho}$ in the phase-space \cite{Gilmore}. The Wigner function of the reference Gaussian state $\hat {\rho}_{G}$ experiences the same deformation in phase space as that of $\hat{\rho}$, and there ore one may conclude that $\hat{\rho}_{G}$ experiences the same Gaussian operation $U_{G}$ as $\hat{\rho}$. Then, using the invariance of trace under cyclic permutations, we have \begin{align} &  \mathrm{Tr}\left(  U_{G} \hat{\rho} U_{G}^{\dagger} \; U_{G}\hat{\rho}_{G}U_{G}^{\dagger}\right)  =\mathrm{Tr}\left(  \hat{\rho}\hat{\rho}_{G}\right) \nonumber\\&  \mathrm{Tr}\left(  U_{G}\hat{\rho}_{G}U_{G}^{\dagger} \; U_{G}\hat{\rho}_{G}U_{G}^{\dagger}\right)  =\mathrm{Tr}\left(  \hat{\rho}_{G}^{2}\right)  ,\end{align} so we conclude that $g$ is invariant under Gaussian operations.

\textit{(ii)} The Gaussianity is a bounded quantity, that is, $2/e\leq g\leq2$. For Gaussian states $g=1$, while the converse is not necessarily true.
 
\textit{Proof.} Having identified in this work the extremum values for $g$ as a function of the uncertainty $\alpha$, it is straightforward to derive the lower and upper bounds of the interval $\left[  2/e,2\right]  $, which correspond to the limit $\alpha\rightarrow\infty$. The state which realizes the lower bound (up to Gaussian transformations) is the number state $\left\vert n\right\rangle $ as $n\rightarrow\infty$, while the upper bound is saturated by the state \begin{equation} \hat{\rho}_{\max}=(1-r)\left\vert 0\right\rangle \left\langle 0\right\vert +r \left\vert n\right\rangle \left\langle n\right\vert\end{equation} with $r\rightarrow0$ and $n\rightarrow\infty$. It is also straightforward to prove that if a state is Gaussian, then $g=1$. On the other hand, $g$ is not constructed on the basis of a mathematical distance, unlike the measures of non-Gaussianity of refs.~\cite{Genoni,Genoni2,Simon}. As a consequence, the inverse statement is not necessarily true and non-Gaussian states might exist possessing a degree of Gaussianity equal to $1$.

\textit{(iii)} The Gaussianity provides necessary criterion for strict positivity of the Wigner function of a state.

\textit{Proof.} In a previous work \cite{Positivity}, we have derived bounds on the trace overlap $\mathrm{Tr}\left(  \hat{\rho}\hat{\rho}_{G}\right)  $ for states with strictly positive Wigner function. These bounds have been derived partially analytically and can be easily translated into bounds on the Gaussianity $g$. Based on the formulas derived in \cite{Positivity}, we can prove that a state with strictly positive Wigner function and of uncertainty $\alpha$ is restricted to have a Gaussianity $g$ that lies in the interval $\left[  g_{\min},g_{\max}\right]  $, where \begin{align} g_{\min} &=0.0095\alpha+0.62+0.711/\alpha-0.333/\alpha^{2}\nonumber\\ g_{\max}  &  =\sqrt{\frac{2}{1+\frac{1}{\alpha^{2}}}}. \label{pos}\end{align} \begin{figure}[t] {\centering{\includegraphics*[width=0.3\textwidth] {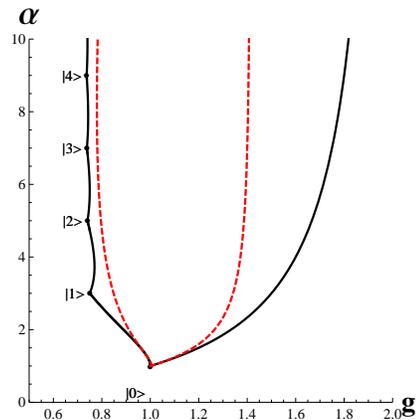}}}\caption{The red dotted line represents Eqs.(\ref{pos}). The black solid line represents the bounds on Gaussianity $g$ derived in the main text. No quantum state with completely positive Wigner function can exist in the area outside the two dotted lines. No quantum state can exist outside the two solid lines.} \label{FIG1}\end{figure}

It should be noted that the bounds on the Gaussianity $g$ given by our Gaussianity-bounded uncertainty relation provide a necessary criterion for a quantum state to be physical. The classical bivariate distributions which achieve the minimum degree of Gaussianity among all continuous positive-definite distributions (classical and quantum quasi-distributions) of the same covariance matrix have been identified in a previous work \cite{Hudson}. Interestingly, one can always find a classical distribution that possess a smaller $g$ than what is allowed for all quantum states. This indicates that the positivity of the density matrix that is imposed to derive the Gaussianity-bounded uncertainty relation is more restrictive than the positivity and continuity of a distribution that is imposed on the proof of \cite{Hudson}. In Fig.\ref{FIG1} we represent the Eqs.(\ref{pos}) by dotted red lines, while the ultimate bounds on $g$, as these derived in the main text, by solid black lines. For a state outside the dotted lines we know that its Wigner fuction has necessarily negative parts. Outside the black lines no quantum states can exist.

To obtain more intuition about the quantity $g$, it is instructive to explicitly express it in terms of the moments of the Wigner quasi-probability distribution. Let us restrict ourselves to a state $\hat{\rho}$ with a covariance matrix $\mathbf{\gamma}=\frac{\hbar}{2}\,\mathrm{diag} (\alpha,\alpha)$, since by linear canonical transformation all states can be reduced to this form and we have proven that $g$ remains invariant along this transformation. In the Wigner representation, the corresponding reference Gaussian state $\hat{\rho}_{G}$ of such a state is phase-invariant; hence, it can be expressed as \begin{equation} W_{G}\left(  r\right)  =\frac{1}{\pi\alpha}e^{-r^{2}/\alpha},\qquad r=\sqrt{x^{2}+p^{2}}. \label{Wgg} \end{equation} In contrast, the state $\hat{\rho}$ itself may possess an angular-dependent Wigner function $W\left(  r,\varphi\right)  $. The trace overlap between $\hat{\rho}$ and $\hat{\rho}_{G}$ takes the following form in the Wigner representation \begin{equation} \mathrm{Tr}\left(  \hat{\rho}\hat{\rho}_{G}\right)  =2\pi{\displaystyle\iint}W\left(  r,\varphi\right) W_{G}\left(  r\right)  r\mathrm{d}r\mathrm{d} \varphi\label{Tr} \end{equation} while we have $g=\alpha\mathrm{Tr}\left(  \hat{\rho}\hat{\rho}_{G}\right)  $.

Now let us average the phase of the Wigner function $W\left(  r,\varphi\right)  $ in order to construct a new state $\hat{\rho}_{s}$ with phase-invariant Wigner function $W_{s}\left(  r\right)  $, \begin{equation}W_{s}\left(  r\right)  =\frac{1}{2\pi}\int W\left(  r,\varphi\right)d\varphi. \label{Wss}\end{equation} Employing Eq.(\ref{Wss}), the trace overlap Eq.~(\ref{Tr}) can be re-written as $\mathrm{Tr}\left(  \hat{\rho}\hat{\rho}_{G}\right)  =4\pi^{2}\int W_{s}\left( r\right)  W_{G}\left(  r\right)  r\mathrm{d}r$, and by expanding \ $W_{G}\left(  r\right)  $ in Taylor series we arrive to \begin{equation} g=4\pi{\displaystyle\sum\limits_{n=0}^{\infty}}\left(  \frac{\left(-1\right)  ^{n}\left\langle r^{2n+1}\right\rangle }{n!\left(  \alpha\right)^{n}}\right)  . \label{gexp}\end{equation} For a Gaussian bivariate distribution independent of the phase as in Eq.(\ref{Wgg}), a simple expression exists for the radial moments \ $\left\langle r^{2n+1}\right\rangle _{G}=\left(  \alpha\right)  ^{n}\Gamma\left(  n+1\right)  /2\pi$. Since by definition $\left\langle r^{3}\right\rangle =$\ $\left\langle r^{3}\right\rangle _{G}$, we conclude that the Gaussianity $g$ accounts for the difference of the odd ($\geq5$) radial moments of the Wigner function of the phase-averaged state \ $\hat{\rho}_{s}$ as compared to those of the reference Gaussian state.

We can use now the phase averaged state \ $\hat{\rho}_{s}$ introduced in Eq.~(\ref{Wss}) to prove that in our search for quantum states which extremize $g$ while possessing a covariance matrix of the form $\mathbf{\gamma}=\frac{\hbar}{2}\,\mathrm{diag}(\alpha,\alpha)$, we are allowed to restrict to mixtures of number states. By substitution of Eq.~(\ref{Wss}) into Eq.~(\ref{Tr}), it is evident that \begin{equation} \mathrm{Tr}\left(  \hat{\rho}\hat{\rho}_{G}\right) =\mathrm{Tr}\left(\hat{\rho}_{s}\hat{\rho}_{G}\right)  .\end{equation} In addition the reference Gaussian state (and therefore the covariance matrix) of $\hat{\rho}_{s}$ is the same as for $\hat{\rho}$, since phase averaging cannot affect the phase-independent Wigner function Eq.~(\ref{Wgg}). From this, one may conclude that \ $g$ is the same for $\hat{\rho}$ and $\hat{\rho}_{s}$ and therefore, invariant under the phase randomization procedure for states with covariance matrix of the form $\mathbf{\gamma}=\frac{\hbar}{2}\,\mathrm{diag}(\alpha,\alpha)$. Therefore, we may with no loss of generality confine our search to states with phase-independent Wigner function, which can be always expressed as mixtures of number states, i.e., $\hat{\rho}=\sum c_{n}\left\vert n\right\rangle \left\langle n\right\vert $.

\begin{figure}[t]{\centering{\includegraphics*[width=0.3\textwidth] {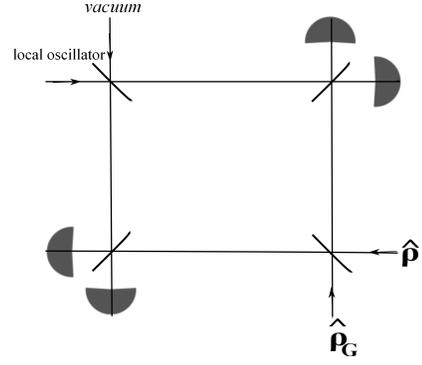}}}\caption{The modified eight-port homodyne scheme which can be used to estimate Gaussianity $g$. In the usual scheme which used to estimate the $Q$-function of a state $\hat{\rho}$, the vacuum state $\left\vert 0\right\rangle $ is fed in the down-right beam splitter instead of $\hat{\rho}_{G}$. } \label{FIG2}\end{figure}

Finally, as explained in the main text, there are possible experimental means for a direct estimation of the Gaussianity $g$ of a state $\hat{\rho}$, without going through a full state tomography procedure. The trace overlap between $\hat{\rho}$ and $\hat{\rho}_{G}$ can be estimated without the need for a full state tomography, using the modified eight-port homodyne detection scheme \cite{Arthur} \ shown in Fig.\ref{FIG2}. However, this would first require performing homodyne measurements on $\hat{\rho}$ for the identification of its covariance and the reconstruction of the reference Gaussian state $\hat{\rho}_{G}$. As a consequence, the whole procedure requires multiple copies of $\hat{\rho}$ just as does a full state tomography procedure, but on the other hand it avoids the instabilities which appear in the reconstruction of the Wigner function from experimental data. Other scenarios for measuring $g$ without constructing $\hat{\rho}_{G}$ from $\hat{\rho}$ may exist, and it is a subject that deserves further investigation.

\section{Analytic expression for the Gaussianity-bounded uncertainty relation}

From the general solution for the Gaussianity-bounded MSs as derived in the main text, we notice that for the interval $\frac{3}{4}<g\leq1$ the uncertainty $\alpha$ is saturated by mixtures of $\left\vert 0\right\rangle $ and $\left\vert 1\right\rangle $. \ The dependence of the minimum value of the uncertainty on Gaussianity degree \ can be easily derived by solving the equation \begin{equation} 4\alpha\frac{\left(  \alpha-1\right)  ^{n}}{\left(  \alpha+1\right)  ^{2+n} }\left(  1+n\right)  =g \label{poly2} \end{equation} for $\alpha$ when $n=0$. One arrives to the following inequality \[ \alpha\geq\frac{2+2\sqrt{1-g}-g}{g},\qquad\mathrm{if~}\frac{3}{4}<g\leq1. \]

For $g>1$, the MSs have been proven in the main text \ to comprise the following set, \begin{equation}\hat{\rho}=\left(  1-r\right)  \left\vert 0\right\rangle \left\langle 0\right\vert +r\left\vert n\right\rangle \left\langle n\right\vert \label{romax2} \end{equation} in the limit $n\rightarrow\infty$, $r\rightarrow0$, and while $\alpha =(1-r)+r(2n+1)$ is kept finite. Employing the identity \begin{equation} g=(1-r)\frac{2\alpha}{\left(  \alpha+1\right)  }+r\frac{2\alpha\left( \alpha-1\right)  ^{n}}{\left(  \alpha+1\right)  ^{n+1}} \label{rr22} \end{equation} one can derive the following dependece of the lower limit of uncertainty as a function of Gaussianity,\[\alpha\geq\frac{g}{2-g},\qquad\mathrm{if~}g>1.\] Thus, ignoring the remaining tiny interval $\frac{2}{\mathrm{e}}<g\leq\frac{3}{4}$ (note $\frac{3}{4}-\frac{2}{\mathrm{e}}=0.014$), we can summarize the Gaussianity-bounded uncertainty relation as \[\alpha\geq\left\{\begin{array}[c]{c}\frac{2+2\sqrt{1-g}-g}{g},\qquad\mathrm{if~}\frac{3}{4}<g\leq1\\\frac{g}{2-g},\qquad\mathrm{if~}g>1.\end{array}\right.\]

\section{Generalization to non-linear constraints}

The method we have developed here in order to re-derive the SR uncertainty relation and derive the Gaussianity-bounded uncertainty relation, is applicable not only in the case where all constraints are linear in the density matrix elements of $\hat{\rho}$, i.e., of the form $\mathrm{Tr}\left( B\hat{\rho}\right)  $, but also in the case where we have non-linear constraints such as the purity $\mathrm{Tr}\left(  \hat{\rho}^{2}\right)  $ or the von Neumann entropy $-\mathrm{Tr}\left(  \hat{\rho}\ln\hat{\rho}\right)$. It is not difficult to show that in the presence of non-linear constraints, the necessary condition on the existence of degeneracies in the spectrum of the Hermitian operator constructed from the constraints is lifted. The necessary condition is, in this case, that every eigenvector $\left\vert \psi_{n}\right\rangle $ of the density matrix of the solution $\hat{\rho}=\sum\left\vert \psi_{n}\right\rangle \left\langle \psi_{n}\right\vert $ should be an eigenvector with positive eigenvalue of a Hermitian operator derived in a similar way as in the main paper \begin{equation} \left(  H-\hat{\rho}\right)  \left\vert \psi_{n}\right\rangle =0. \label{gc}\end{equation} Moreover, as it is dictated by the above condition Eq.~(\ref{gc}), the eigenvalues are the mixing amplitudes of the eigenvectors \begin{equation} H\left\vert \psi_{n}\right\rangle =\left\langle \psi_{n}\right\vert \left. \psi_{n}\right\rangle \left\vert \psi_{n}\right\rangle =c_{n}\left\vert\psi_{n}\right\rangle .\end{equation} In a further work \cite{furtherwork}, we explicitly show via an example how the method develops in this obviously more complicated case and we draw parallels with the method of derivation of purity -bounded uncertainty relations \cite{Man'ko,Dodonov}.




\begin{thebibliography}{99}                                                                                             
\bibitem {Heisenberg}W. Heisenberg, Z. Phys. \textbf{43}, 172 (1927).

\bibitem {Kennard}E. H. Kennard, Z. Phys. \textbf{44}, 326 (1927).

\bibitem {Weyl}H. Weyl, \textit{Theory of groups and quantum
mechanics}, New York: Dutton, pp. 77, 393-394 (1927).

\bibitem {Mandel'shtam}L. I. Mandel'shtam and I. E. Tamm, \textit{
The uncertainty relation time-energy in nonrelativistic quantum
mechanics}, Izv. AN USSR, Seriya Fiziki., vol. 9, no.1/2, pp.
122-128 (1945).

\bibitem {Moyal}J. E. Moyal, Proc. Cambridge Philos. Soc. \textbf{45}, 99 (1949).

\bibitem {Stoler}D. Stoler and S. Newman, Phys. Lett. A \textbf{38}, 433 (1972).

\bibitem {Man'ko}V. V. Dodonov and V. I. Man'ko, in: \textit{ Invariants and Evolution
of Nonstationary Quantum Systems}, Proc. Lebedev Physics Institute, Vol. 183,
edited by M. A. Markov (Nova Science, Commack, NY, 1989), pp. 3-101.

\bibitem {Schrodinger}E. Schr\"{o}dinger, Sitzungsber. Preuss. Akad. Wiss. 
\textbf{14}, 296 (1930)

\bibitem {Robertson}H. P. Robertson, Phys. Rev. \textbf{35,} 667A (1930); ibid.  \textbf{46,} 794 (1934)

\bibitem {Kurmyshev}V. V. Dodonov, E. V. Kurmyshev and V. I. Man'ko, Phys.
Lett. A \textbf{79}, 150 (1980).

\bibitem {Scully} M. O. Scully and M. S. Zubairy, \textit{Quantum optics}, (Cambridge University Press, Cambridge, 1997). 

\bibitem{Hammerer} K. Hammerer, A. Sorensen, and E. S. Polzik, Rev. Mod. Phys. \textbf{82}, 1041 (2010).

\bibitem{Aspelmeyer} S. Groeblacher, K. Hammerer, M. R. Vanner, M. Aspelmeyer, Nature \textbf{460}, 724 (2009).

\bibitem{Chiorescu} I. Chiorescu, Y. Nakamura, C. J. P. M. Harmans and J. E. Mooij, Science \textbf{299}, 1869 (2003).


\bibitem {Dodonov}V.~V.~Dodonov, J.~Opt.~B: Quantum Semiclass. Opt. \textbf{4}, S98-S108 (2002).

\bibitem {Bastiaans}M. J. Bastiaans, J. Opt. Soc. Am. A, 1243 (1986).

\bibitem {Sudarshan}E. C. G. Sudarshan, C. B. Chiu and G. Bhamathi, Phys. Rev.
A \textbf{52}, 43 (1995).

\bibitem {Simon2}R. Simon, N. Mukunda and B. Dutta, Phys. Rev. A \textbf{49},
1567 (1994).

\bibitem {Ponomarenko}S. A. Ponomarenko and E. Wolf, Phys. Rev. A \textbf{63,}
062106 (2002); G. S. Agarwal and S. A. Ponomarenko, Phys. Rev. A \textbf{67,}
032103 (2003).


\bibitem {Polzik}J. S. Neergaard-Nielsen, B. M. Nielsen, C. Hettich, K. Molmer, and E. S. Polzik, Phys. Rev. Lett. \textbf{97}, 083604 (2006).

\bibitem {Grangier2}A. Ourjoumtsev, H. Jeong, R. Tualle-Brouri and P. Grangier, Nature \textbf{448}, 784 (2007).

\bibitem{Parigi07} V. Parigi, A. Zavatta, M. Kim, and M. Bellini, Science \textbf{317}, 1890 (2007).

\bibitem {Furusawa2}H. Takahashi, K. Wakui, S. Suzuki, M. Takeoka, K. Hayasaka, A. Furusawa, and M. Sasaki, Phys. Rev. Lett. \textbf{101}, 233605 (2008).

\bibitem{Zavatta09} A. Zavatta, V. Parigi, M. S. Kim, H. Jeong, and M. Bellini, Phys. Rev. Lett. \textbf{103}, 140406 (2009).

\bibitem{Bimbard10} E. Bimbard, N- Jain, A. MacRae, and A. I. Lvovsky, Nat. Photon. \textbf{4}, 243 (2010).


\bibitem{NoGoPurif1} J. Eisert, S. Scheel, and M. B. Plenio, Phys. Rev. Lett.
\textbf{89,} 137903 (2002).

\bibitem{NoGoPurif2} J. Fiurasek, Phys. Rev. Lett. \textbf{89,} 137904 (2002).

\bibitem{NoGoPurif3} G. Giedke and J. I. Cirac, Phys. Rev. A \textbf{66,}
032316 (2002).

\bibitem{NoGoCorr} J. Niset, J. Fiurasek, and N. J. Cerf, Phys. Rev. Lett.
\textbf{102,} 120501 (2009).

\bibitem {NoGoCommit}L. Magnin, F. Magniez, A. Leverrier, and N. J. Cerf,
Phys. Rev. A \textbf{81,} 010302(R) (2010).


\bibitem{Braunstein} S. L. Braunstein, C. M. Caves, and G. J. Milburn, Ann. Phys. \textbf{247}, 135 (1996).

\bibitem {Genoni}M.~G.~Genoni, M.~G.~A.~Paris, and K.~Banaszek, Phys.~Rev.~A
\textbf{76}, 042327 (2007).

\bibitem {Genoni2}M. G. Genoni, M. G. A. Paris and K. Banaszek, Phys. Rev. A
\textbf{78}, 060303(R) (2008); M. G. Genoni and M. G. A. Paris, Phys. Rev. A
\textbf{82}, 052341 (2010).

\bibitem {Simon}J. Solomon Ivan, M. Sanjay Kumar, and R. Simon, Quantum Inf.
Process. \textbf{11}, 853 (2012).

\bibitem{Hudson}A. Mandilara, E. Karpov, and N. J. Cerf,
Phys. Rev. A \textbf{79}, 062302 (2009).

\bibitem {Positivity}A. Mandilara, E. Karpov and N. J. Cerf, J. Phys.: Conf.
Ser. \textbf{254}, 012011 (2010).

\bibitem {Cerf} N. J. Cerf, G. Leuchs, and E. S. Polzik (eds.), \textit{Quantum information with Continuous Variables of Atoms and Light}, (Imperial College Press, London, 2007).

\bibitem {Gilmore}W.-M. Zhang, D. H. Feng and R. Gilmore, Rev. Mod. Phys. 62, 867 (1990).

\bibitem {Arthur}E. Arthurs and J. Kelly, Bell System Tech. J. \textbf{44}, 725 (1965).

\bibitem {furtherwork} A. Mandilara, E. Karpov and N. J. Cerf, in preparation.




\end{thebibliography}
\end{document}